# Low Latency Video Denoising for Online Conferencing Using CNN Architectures

Altanai Bisht, Ana Carolina de Souza Mendes, Justin David Thoreson II, Shadrokh Samavi

abisht@seattleu.edu, adesouzamendes@seattleu.edu, thoresonjust@seattleu.edu, samavishadro@seattleu.edu

*Department of Computer Science,* Seattle University, Seattle, WA 98122, USA

*Abstract*—In this paper, we propose a pipeline for real-time video denoising with low runtime cost and high perceptual quality. The vast majority of denoising studies focus on image denoising. However, a minority of research works focusing on video denoising do so with higher performance costs to obtain higher quality while maintaining temporal coherence. The approach we introduce in this paper leverages the advantages of both image and video-denoising architectures. Our pipeline first denoises the keyframes or one-fifth of the frames using HI-GAN blind image denoising architecture. Then, the remaining four-fifths of the noisy frames and the denoised keyframe data are fed into the FastDVDnet video denoising model. The final output is rendered in the user's display in real-time. The combination of these low-latency neural network architectures produces real-time denoising with high perceptual quality with applications in video conferencing and other real-time media streaming systems. A custom noise detector analyzer provides real-time feedback to adapt the weights and improve the models' output.

## I. INTRODUCTION

The adoption of real-time video conferences accelerated due to the COVID-19 pandemic, lockdown, and the shift towards remote learning, work from home, and telemedicine. Corporations, schools, and families use video conferencing applications daily for communication. This lifestyle change continues to be utilized, with remote work and education being more accepted as an effective medium. However, the user experience in video conferences is tightly tied to the playback quality, which can be defined as having synchronized audio and video, lossless media stream, and homogenous content transmitted to all participants. There are existing solutions to filter audio noise but improving real-time video transmission quality is challenging.

Video noise has various root causes, such as webcam or hardware issues, laptop, and mobile processing cycles, environment issues like lighting or motion, network transmission delays due to congestion, data corruption, and packet-loss during transmission. The challenges of denoising videos stem from their high amount of data, reluctance to retransmit lost keyframes, the need to maintain temporal coherence between each denoised frame, and the lack of performance in high-quality media streaming solutions. In addition, real-time video denoising introduces the extra challenge of denoising each frame with a low runtime cost.

Reference [1] compares image and video denoising. It contrasts that image denoising is a popular field of research in comparison to video denoising, which lacks exploration in literature. Similarly, this paper explores image-denoising techniques to hypothesize an architecture that offers high-quality image-denoising when integrated with video streaming pipelines. In video denoising, one model [1] was analyzed due to its low runtime results compared with other algorithms discussed in that paper.

The rest of this paper is organized as follows. In Section 2, the literature review done for this research is explained in detail. Section 3 is dedicated to experimental results from the related work introduced in Section 2, closed with a discussion about the results. Section 4 explains the proposed method for introducing video denoising to the video conferencing pipeline. Section 5 covers the pipeline architecture, and in Section 6, concluding remarks are presented.

## II. LITERATURE REVIEW

### A. General Image Denoising Techniques

*1) Using Hardware*

One popular image-denoising method is to use specialized hardware, such as FPGA devices. This solution is attractive for its high performance, Peak Signal-to-Noise Ratio (PSNR) results, and preserving image details, as proposed by [2]. Although studies present satisfactory results, hardware solutions require the hardware device to be embedded, which also means an increase in financial cost overhead. To utilize the same architecture for various videoconferencing applications, specialized hardware solutions that needed embedding, such as FPGA proposed by [2], were quickly ruled out as a viable alternative to this research.

In search of scalable hardware options, existing processing options in personal computers (CPUs and GPUs) and how they are currently leveraged in existing denoising solutions were explored to define the proposed architecture. Results from [1] and [3] are summarized in table 1 below.

CPU AND GPU USAGE IN DENOISING TECHNIQUES

| Device | Total Methods | Average Run Time (s) |
|--------|---------------|----------------------|
| CPU    | 6             | 177.33               |
| GPU    | 6             | 0.90                 |

Table 1: Summary of total denoising techniques, the device used for processing, and the average run time. Data were collected from [1] and [3].

Based on the average run times, denoising architectures that utilize GPUs for processing achieved high-quality results while maintaining a low run time cost. This study's proposed method is also compatible with being processed using GPU-based toolkits such as CUDA to achieve faster real-time denoising results.

*2) Using Artificial Intelligence*

As advancements within technological ground have occurred, particularly the breakthroughs within artificial intelligence and machine learning, image-denoising techniques have made significant progress insofar that many specific methods share similar architectures and generalized, high-level approaches. Current denoising practices involve but are not limited to, the adoption of spatial domain methods, transform techniques, and convolutional neural networks (CNNs) [4]. CNNs, in particular, are more heavily considered nowadays due to the success of their denoising quality; the adoption of deep CNNs (DCNNs) [3, 4] and Autoencoders [5] show substantial developments in image denoising.

Though CNNs are generalized architectures, they themselves can be applied as components to other generalized methods, which can, in turn, be applied to specific use case implementations. For instance, another generalized technique used in image denoising is the usage of generative adversarial networks (GANs), which are comprised of two main DCNNs: a generator and a discriminator. The generator and the discriminator are adversaries competing in a zero-sum game (or Minimax game). The generator produces fake images to fool the discriminator into classifying them as real instead of fake. In contrast, the discriminator seeks to classify all non-generated images as real and all generated images as fake. Both DCNNs are trained together to improve each of their models insofar that there will be a fifty percent chance of classifying any image, non-generated or generated, as real or fake. This characteristic means that the generator is capable of producing quality images that appear just as real as non-generated images. In denoising, the generator can take a noisy image as input and generate a denoised image that is relatively indistinguishable from a non-generated, non-noisy image due to the significant preservation of original features.

Reference [5] presented many variations of CNNs used for image denoising. There are tradeoffs in terms of perceptual and objective qualities, preservation of details such as texture or a lack thereof, and the existence of artifacts. For real-time video conferencing denoising, it is preferred to achieve high perceptual quality while preserving details with low artifacts. A high peak signal-to-noise ratio (PSNR) is preferable for denoising images but incurring low run time cost is valued over high objective quality in real-time applications. With this goal defined, other denoising techniques that meet these use-case requirements were researched.

*B. Hierarchical Generative Adversarial Network (HI-GAN)*

Many specific solutions to image denoising have embraced the achievements of DCNNs and GANs; one specific implementation is HI-GAN [3]. HI-GAN expands the typical GAN model by employing two additional generators; thus, a hierarchical image-denoising structure is formed in a successful attempt to attain high denoising quality via the preservation of original features and elevated PSNRs [3].

Like GAN, the first generator is trained with the single discriminator in the same zero-sum game described before [3]. Though the single generator model has great preservation of original features, the resulting PSNR values tend to be lower than desired, therefore warranting the implementation of a second generator [3]. The second generator receives the first generator's output image as input and processes the image to increase the PSNR value [3]. The resulting PSNR values produced by the second generator have significant improvements. However, the images lose some original features that were preserved by the first generator [3]. Thus, a third generator takes the two previous generators' output images as input and generates a final denoised image with significant preservation of original features and a high PSNR value [3].

*C. Fast Deep Video Denoising Network (FastDVDnet)*

The aforementioned work in this section is catered toward handling the denoising of individual images, but the scope of our work involves real-time video. Video denoising sees significantly less attention than image denoising; thus, existing video denoising techniques are fewer. Luckily, recent research into video denoising has resulted in successful experimentation developments.

Non-blocking methods such as VBM4D[8] can do video denoising through Separable 4-D Nonlocal temporal transformations. However, CNN models such as VNLNet[9] and FastDVDnet[6]. have been shown to outperform these. FastDVDnet is a video-denoising algorithm that incorporates CNNs in the form of U-Net architecture [1]. As illustrated in Figure 1, FastDVDnet receives five contiguous frames as input, with the center frame being the keyframe to denoise [1]. The algorithm then runs two steps that are quite similar to one another [1]. The first step involves three denoising blocks [1].

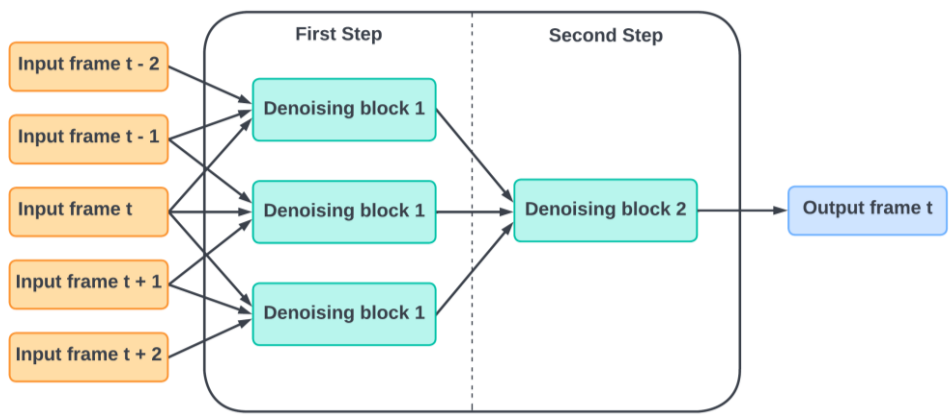

Fig. 1. High level diagram of FastDVDnet architecture. [1]

Out of the five contiguous frames, the first three frames, the middle three frames, and the last three frames are passed to the first denoising block, the middle denoising block, and the last denoising block, respectively [1]. Each denoising block will produce a denoised frame based on the three input frames received. Then, step two occurs, which involves only one denoising block [1]. This denoising block produces a denoised frame in the same way as the previous three denoising blocks. However, instead of taking three contiguous video frames as input, it takes the three output frames from step one [1]. The output of this denoising block is the final denoised frame [1].

### III. EXPERIMENTAL RESULTS

#### A. Hierarchical Generative Adversarial Network (HI-GAN)

The HI-GAN algorithm was tested using six different datasets: DND, SIDD, NAM, NC12, FMD, and EFMD [3]. This algorithm denoised images with higher PSNR and SSIM than all other related algorithms. Two of the results from the paper are displayed in Table 2. It is interesting to note that HI-GAN performed better than all other methods with blind and non-blind denoising.

Table 2: HI-GAN quantitative experiment results [3]. a) Quantitative results on the Nam dataset. b) Quantitative results on the DND dataset.

| Methods | Mode | PSNR (dB) | SSIM |
|---|---|---|---|
| CDnCNN-B | Blind | 37.67 | 0.9379 |
| WNNM | Non-blind | 38.79 | 0.9488 |
| MCWNNM | Blind | 39.16 | 0.9623 |
| CBDNet | Blind | 39.36 | 0.9649 |
| TWSC | Blind | 40.05 | 0.9714 |
| UNet-D | Non-blind | 40.23 | 0.9709 |
| UNet-ND | Non-blind | 40.65 | 0.9735 |
| **HI-GAN** | **Non-blind** | **40.85** | **0.9741** |

a)

| Methods | Mode | PSNR (dB) | SSIM |
|---|---|---|---|
| CDnCNN-B | Blind | 32.43 | 0.7900 |
| TNRD | Non-blind | 33.65 | 0.8306 |
| FFDNet | Non-blind | 34.40 | 0.8474 |
| WNNM | Non-blind | 34.67 | 0.8648 |
| GCBD | Blind | 35.58 | 0.9217 |
| CIMM | Non-blind | 36.04 | 0.9136 |
| MCWNNM | Blind | 37.38 | 0.9294 |
| TWSC | Blind | 37.94 | 0.9403 |
| CBDNet | Blind | 38.06 | 0.9421 |
| UNet-D | Blind | 38.59 | 0.9467 |
| UNet-ND | Blind | 38.70 | 0.9483 |
| **HI-GAN** | **Blind** | **39.37** | **0.9542** |

b)

Regarding run time cost, HI-GAN performed better than algorithms running on CPU. HI-GAN ran slower than other methods on the GPU, namely CBDNet, and slower than FFDNet. However, the advantage of HI-GAN is that it limits the running time [3], which is ideal for real-time applications.

Moreover, we analyzed the tradeoff in this HI-GAN's runtime cost and denoising quality and found it feasible for our proposed video conferencing media pipeline. The run time results are displayed in Table 3.

Table 2: Running time of HI-GAN and competing methods [3].

| Methods | Device | Run time (s) |
|---|---|---|
| CDnCNN-B | CPU | 79.5 |
| TNRD | CPU | 5.2 |
| FFDNet | GPU | 0.01 |
| CBDNet | GPU | 0.1 |
| MCWNNM | CPU | 208.1 |
| TWSC | CPU | 195.2 |
| **HI-GAN** | **GPU** | **0.2** |

#### B. Fast Deep Video Denoising Network (FastDVDnet)

The FastDVDnet algorithm was tested using two different datasets: DAVIS-test and Set8 [1]. The videos were denoised by both FastDVDnet and DVDnet with high temporal coherence, which is desired for good-quality videos. Overall, the PSNR values for various noise ratios were higher or on par with other denoising algorithms. The results are summarized in Figure 2.

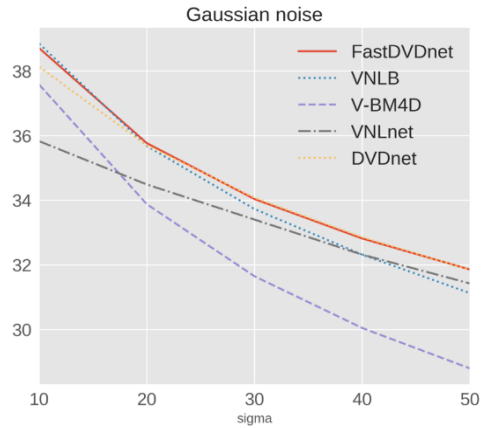

Fig. 2. FastDVDnet experiment results compared with related methods [6].

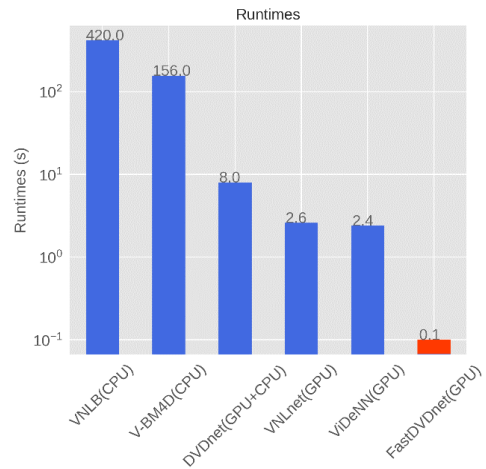

Fig. 3. FastDVDnet running time compared with related methods [6].

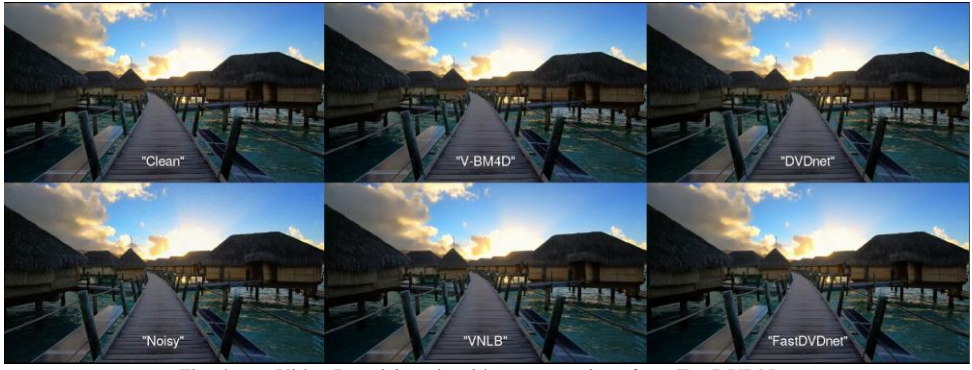

Fig. 4. Video Denoising algorithms comparison from FastDVDNet.

FastDVDnet performed better than all competing methods running on CPUs or GPUs in terms of run time cost. The run time results are displayed in Figure 3. Additionally, it bears a smaller memory footprint. Figure 4 depicts the video denoising quality obtained from FastDVDnet, its percursor DVDnet, V-BM4D[8], and VNLB[9].

## C. Discussion

The HI-GAN architecture achieved all the goals of this paper. The denoised images presented both high PSNR and perceptive quality while maintaining texture details. The runtime cost was low, on par with other state-of-the-art algorithms. The next challenge was verifying if HI-GAN applies to video denoising. To better understand video quality and denoising, the [1] was studied and provided clarity on the topic.

The FastDVDnet was the best-performing algorithm compared to other video-denoising techniques and on par with image-denoising solutions. In addition, the algorithm successfully maintained details from texts, which is ideal for video conferences where participants share materials during presentations.

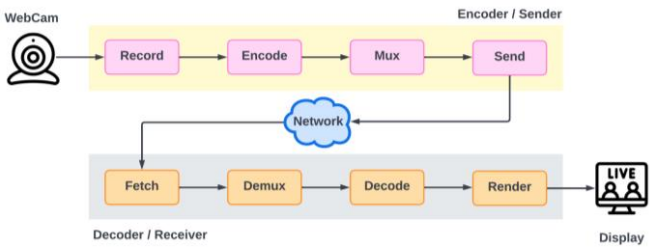

Fig. 5. The media pipeline for video conferencing applications.

## IV. PROPOSED METHOD

### A. The Media Pipeline

The RTC (RealTime Media Connection) pipeline between the sender's media capture and the receiver's playback needs to be elaborated to understand where the video denoising processing should occur. The rest of this paper will address the RTC pipeline as a media pipeline. In Figure 5, the media pipeline is presented, and the two main parts of the media pipeline can be addressed as the sender's and receiver's sides. First, the sender obtained media through the system hardware, raw media data. Then the sender's encoder module compresses it to a format suitable to send over the network. This is often using the RTP (Real Time Protocol) format. These encoded packets traverse the network and may pass through congested bottlenecks or experience packet loss. Next, the receiver's side obtains these packets and populates them in the buffer for further processing. Finally, the decoder module decomposes the data and decodes it to render it on the user's display. At this point, the reconstructed frame in the video may experience degradation because of packet loss during transmission, lossy compression codecs, or even bad lighting conditions around the sender's webcam. Our proposed architecture aims to denoise the received keyframes and the video before it is rendered but after being decoded by the video conferencing application's algorithm. This post-processing module is situated between the decode and render components in the media pipeline, as shown in Figure 6.

## V. PIPELINE ARCHITECTURE

### A. High-level architecture overview

As alluded to in the previous section, our architecture resides between the decoding and rendering portions of the video conferencing media pipeline. After decoding the video frames, the noise detector will assess each keyframe for noise. If the noise distribution within a keyframe is significant, the keyframe is then sent to an image denoiser. The denoising algorithm should have a relatively fast runtime with acceptable denoising quality for this portion. Therefore, we consider HI-GAN and CBDNet as the image denoiser models. The next step is to send the denoised keyframe to a video denoiser. Again, we propose using FastDVDnet as the video denoiser. As such, additional frames immediately preceding and succeeding the keyframe will be used as input to the temporal video denoiser. After the video denoising step, the keyframe is ready to move to the render phase of the video conferencing media pipeline. In addition to rendering the denoised keyframe, the frame will be sent to a denoising analyzer, which will evaluate the denoised image and provide feedback to the denoisers such that they adjust their models according to the resulting quality of their denoising efforts.

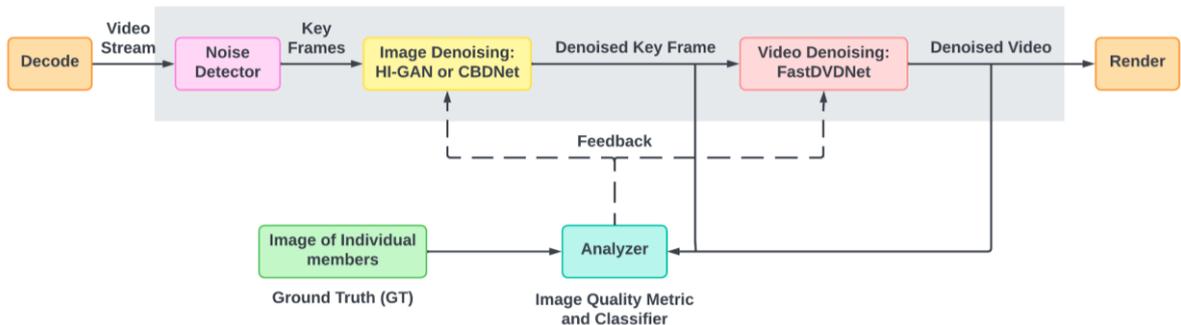

Fig. 6. The proposed method high-level architecture.

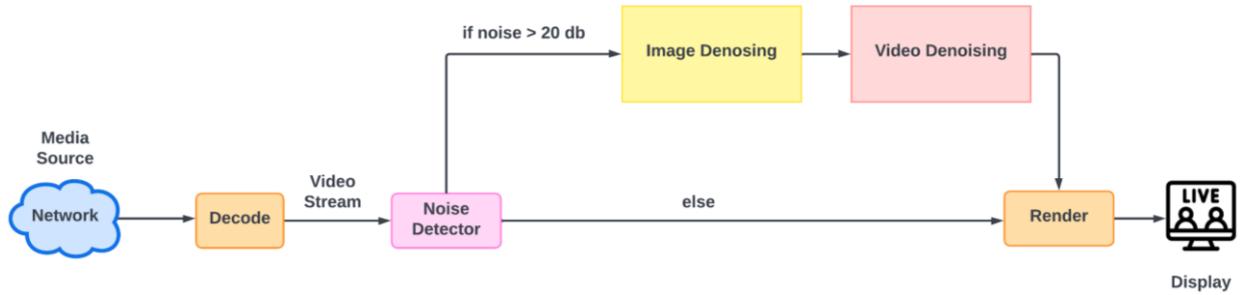

Fig. 7. Noise detector component that evaluates the noise levels(sigma) before deciding to include denoising fork in the pipeline or not.

A typical real-time streaming media pipeline, such as the ones used for video conferencing systems, has codecs. The codec is for compression and decompression, media control units for multiplexing and demultiplexing, network routing optimizations using a bit rate adaptor, feedback, and end-user agents for capture, rendering, and buffering. Our proposed pipeline introduces a noise detector, image denoiser, video denoiser, and denoise analyzer, and the existing real-time streaming media pipeline components. The following sections describe these components in detail.

*B. Noise detector*

The objective of the noise detector is to detect if the noise is above a predefined threshold which is end-user perceivable. The noise is measured in percentage and denoted by sigma. For our experiments, we evaluated noise levels between 20 and 50%. The objective is to measure the noise in the reconstructed frame at the receiver's end of the network from that at the sender's end. Based on the results, the pipeline is forked into two parts. Part one bypasses the denoising pipeline and renders the video keyframes immediately after decoding. Part two of the pipeline is for sending the frames through the image and video denoising and then rejoining the render components for end playback.

Figure 7 presents the high-level overview of the Noise Detector, which decides if the keyframe's noise level is high enough to be sent through the denoising pipeline.

The noise detector estimates the noise levels and categorizes the frame into Gaussian, signal-dependent, and processed noise in image and video signals. Identifying noise and its levels are essential to effectively switching between blurring, neural network-based denoising, or autoencoder-based techniques to yield the fastest acceptable results. The following figure shows the different types of noise identified in real-time media communication, such as Salt and Pepper noise, local variable noise, pixelation due to packet loss or low network bandwidth, and speckle or impulse noise. Video conference systems are prone to packet loss. Multiple factors impact the noise levels in the video frames such as dynamic network condition, codecs profiles or routing algorithms.

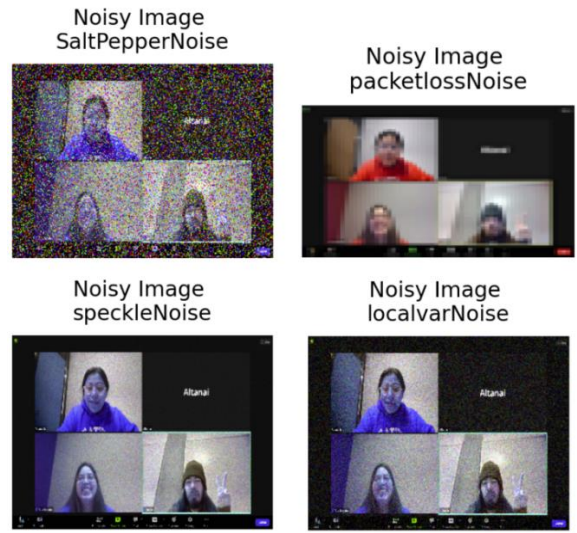

Fig. 9. Some frequently occurring noise types are shown.

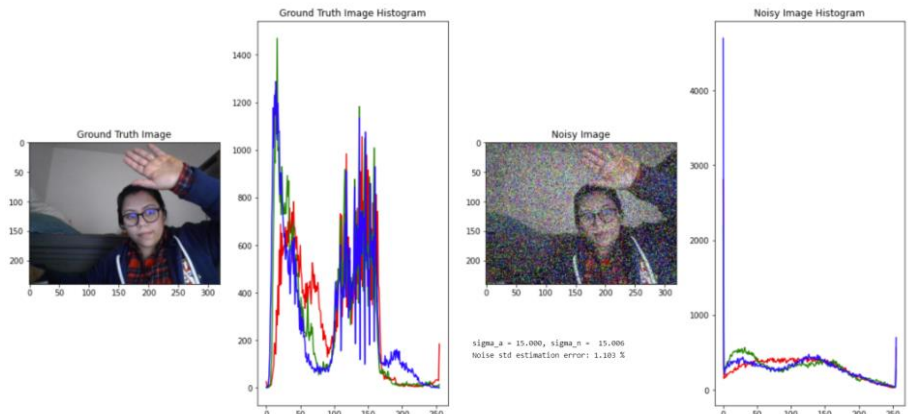

Fig. 8. Analyzing the pixel distribution on few clean-noisy image pairs by plotting histograms as well as calculating sigma.

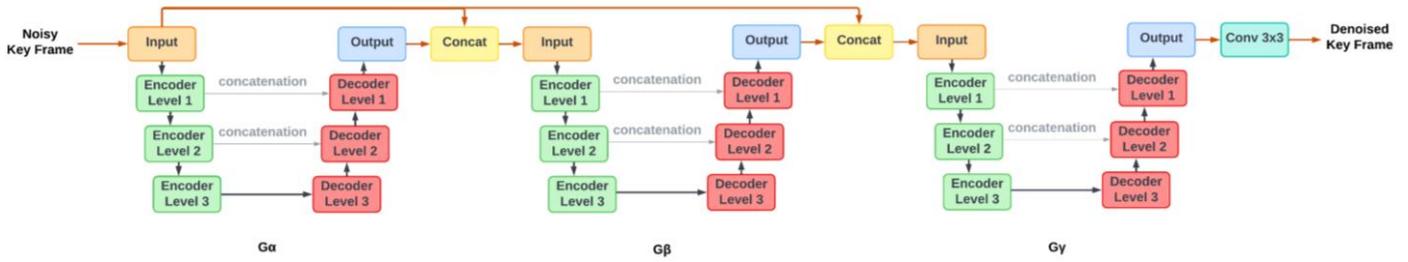

Fig. 10. HI-GAN Generators simplified model.

*C. Image and Video Denoising*

Image Denoiser for the keyframe, also known as I-frame, is hypothesized to help the video denoising process. These denoised keyframes will be the basis for temporal and spatial coherence-based video-denoising algorithms for denoising neighboring frames. The image-denoising model is adapted from HI-GAN and applied to I-frames. In addition, the model is designed to allow skip-connections to fasten the outputs.

We used a video Denoiser using temporal correlation for real-time video frame processing. Successful denoising in the key frame will result in video denoising for the succeeding frames despite the low-quality resolution. This helps keep video playback in real-time without sacrificing bandwidth for buffering for all high-resolution frames. The steps for denoising are outlined in Figure 11.

*D. Denoising Analyzer*

The architecture for the denoising analyzer is shown below in Figure 10. If available, the analyzer evaluates the denoised key frames against existing ground truth images. The ground truth can be either the user's profile image or captured snapshot image from the sender's side before encoding and network traversal. The obtained result is sent as feedback to the image and video denoisers to enable them to adapt and improve their respective weights and biases. Furthermore, illustrated in Figure 12, feedback may be sent to the sender's side of the network. The benefit of this end-to-end feedback is to provide the sender's end with the relevant information to determine whether the receiver's end can denoise the image after network transmission. If the denoising improvements are insignificant, it can signal the encoder to adapt the bitrate or improve the quality. In addition, the encoder/sender may adjust their respective processes to mitigate noise introduction. Some adjustments on the encoder/sender side may include but are not limited to, framerate and resolution.

The analyzer can detect the performance of the denoising system by comparing the delta between the sender and receiver metrics score. The denoising analyzer compares the improvements via metrics such as PSNR, SSIM (Structural Similarity), MS-SSIM (Multi-Scale Structural Similarity), NIQE (Naturalness Image Quality Evaluator), VMAF[7], VIFp (Visual Information Fidelity, pixel domain version) among others. The same is shown in Figure 13.

The neighboring frames are denoised from the denoised keyframes, and the video is reconstructed to be rendered to the receiver. The analyzer compares the faces and objects detected before and after smoothening with the sender side To detect over-smoothening or alteration of the stream. This classification helps to identify whether details are lost in the obtained network or post-processed output.

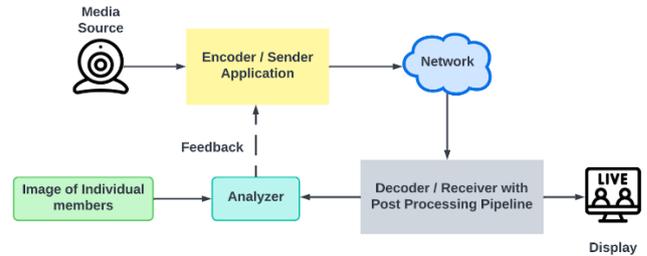

Fig. 12. The denoising analyzer pipeline

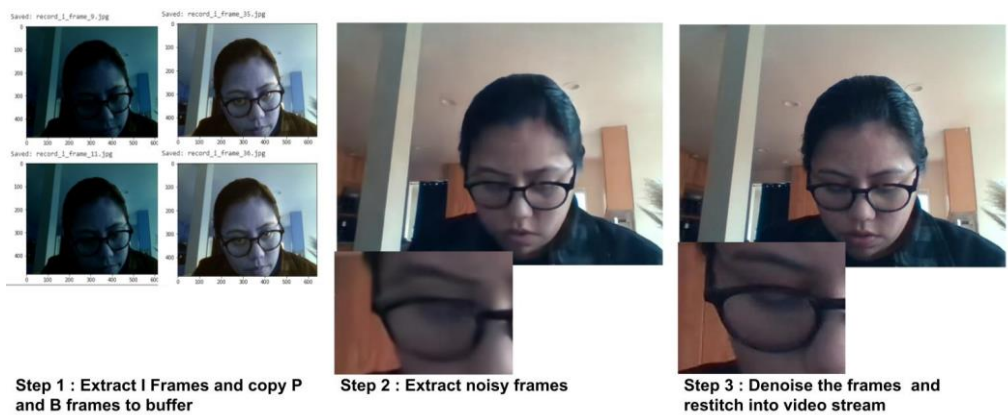

Fig. 11. Steps for processing video key frames in the denoising pipeline.

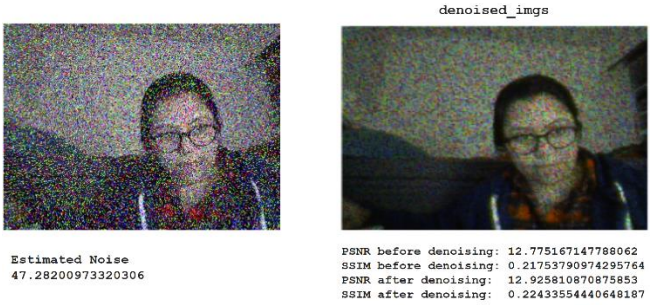

Fig. 13. Analyze the blind noise before and after the media packet traverses the network and evaluate the denoising component's performance.

Our proposed denoising analyzer uses object detection on the sender side media frame and receiver side network traversed, reconstructed and denoised frame to evaluate if details are lost or whether objects are unrecognizable at the far end of the media pipeline.

Figure 14 shows that denoising has smoothened the image at the end of the RTC pipeline. This causes the object detector to fail to detect objects such as "glasses" in the denoised image, which was present in the sender side image. The analyzer's output forms the feedback loop for a real-time denoising system. The metrics calculated by the denoising analyzer will generate a dynamic performance score based on runtime, PSNR, and SSIM improvement. This is immensely useful for the low latency streaming use cases where Time constraint is calculated based on network condition and receiver report on media packets. The feedback is also used as a metric to calculate weights and adapt biases. The analyzer can also share feedback with the sender to convey the end-user experience. The encoder and sender side can gauge the performance of the outgoing stream and adjust the framerate, bitrate, and resolution to ease the denoising process and improve the analyzer score.

## VI. CONCLUSION / FUTURE WORK

### A. Tradeoff Analysis

Through our literature review and experimental analysis, we identified the feasibility of the Autoencoder algorithm in low-latency streaming, such as video conferencing. The Tradeoff Analysis compares the performance of high-compute systems such as GPU-based application programs (CUDA, OpenCL) against low-compute engines such as WebRTC browsers, for example, chromium and Mozilla Firefox. We also factored in the tradeoff between speed and accuracy, as shown in Figure 15. Based on the requirement to maintain low runtime to enable low latency playback, we propose a system that can limit the accuracy to allow noise in non-perceivable levels and monitor SSIM metric improvements to adapt to network conditions.

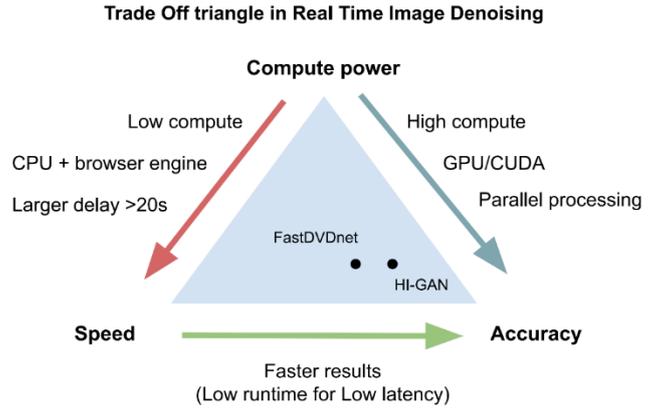

Fig. 15. Tradeoff triangle for real-time denoising.

### B. Experimental Results

Our experiments were conducted in a CPU-based environment using Python 3 Google Compute Engine with 2GB RAM. We evaluated the time to denoise images from media streams obtained through a WebRTC-based video call session on the receiver's end. We selected I-frames from the RTP media stream. We used the noise detector to detect whether the data had noise levels exceeding 20 dB. The identified frames were processed through the denoising pipeline fork. The Denoising analyzer evaluated the resulting performance improvements. Figure 16 depicts the outcome of our analysis. This study found that autoencoder-based GAN models are feasible for real-time denoising with acceptable performance and configurable feedback mechanisms between the sender and receiver.

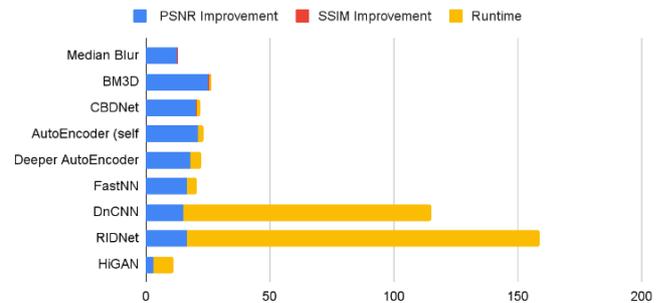

Fig. 16. Performance of Denoising algorithms with metrics of PSNR improvement, SSIM improvement, and runtime.

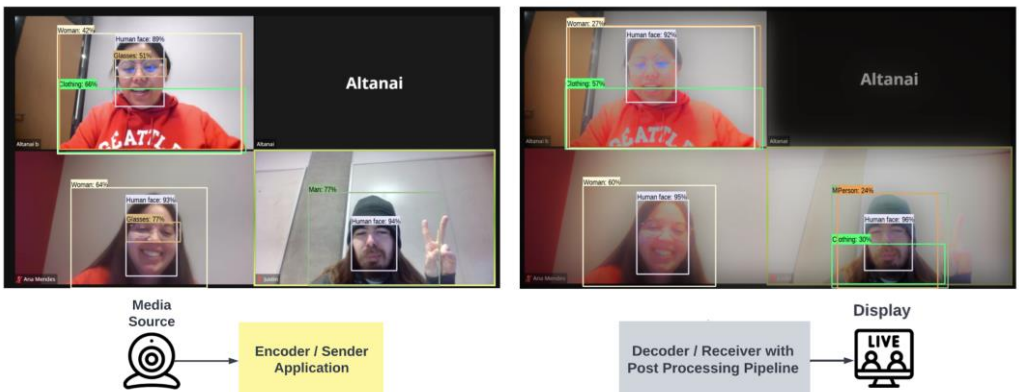

Fig. 14. Object detection at sender and receiver end to evaluate the quality of denoising.

We observed that while ground truth pre-trained denoising algorithms perform faster and more accurately than blind ones, they do not perform well when the participants in the video conference are new or the backgrounds change drastically. Hence blind denoising algorithms are more suited to the use case when the video conference has varying faces and backgrounds. The code supporting our experimentation and analysis can be found within the following GitHub organization:

https://github.com/RealtimeDenoising

### C. Future Scope of Work

The hardware improvement from CPU to GPU is expected to lower the runtime for denoising images and videos significantly. CUDA is a toolkit to enable parallel processing suited to this use case of video denoising [10]. We proposed optimizations to perform targeted denoising by segmenting the image into noisy and non-nosy areas. The identified noisy section of the frame can be processed for denoising in parallel and fed into restitching the result.